\documentclass[aps,prd,floats,floatfix,preprintnumbers,superscriptaddress,nofootinbib]{revtex4-2}

\usepackage{graphicx} 
\usepackage{amsmath}
\usepackage{amssymb}
\usepackage{xcolor}
\usepackage{hyperref}
\usepackage[normalem]{ulem}
\hypersetup{
    colorlinks,
    citecolor=blue,
    filecolor=blue,
    linkcolor=blue,
    urlcolor=blue
}

\newcommand{\be}{\begin{equation}}
\newcommand{\ee}{\end{equation}}

\newcommand{\bea}{\begin{eqnarray}}
\newcommand{\eea}{\end{eqnarray}}

\newcommand{\rmd}{\mathrm{d}}

\begin{document}

\title{Violation of energy conditions and the gravitational radius of the proton}
\author{Adrian Dumitru}
\email{adrian.dumitru@baruch.cuny.edu}
\affiliation{Department of Natural Sciences, Baruch College, CUNY,
17 Lexington Avenue, New York, NY 10010, USA}
\affiliation{The Graduate School and University Center, The City University of New York, 365 Fifth Avenue, New York, NY 10016, USA}

\author{Jorge Noronha}
\email{jn0508@illinois.edu}
\affiliation{Department of Physics and Illinois Center for Advanced Studies of the Universe, University of Illinois Urbana-Champaign, 1110 West Green Street, Urbana, Illinois 61801, USA}


\begin{abstract}
{ The energy–momentum tensor (EMT) of the proton encodes fundamental information about its mass, pressure, and shear distributions. Using recent lattice QCD data for the gravitational form factors, we show that the Breit–frame Wigner EMT  may be of Hawking–Ellis type~IV in the proton’s core. Such EMT violates all pointwise energy conditions and lacks a causal rest frame so that the usual mechanical picture fails at short distances. We define the \emph{gravitational radius}—a new hadronic observable—marking the scale where the EMT becomes ordinary (type I) and the classical interpretation is restored. We also derive from the Averaged Null Energy Condition (ANEC)
non-perturbative, model-independent QFT constraints on gravitational form factors.}
\end{abstract}
\maketitle

\section{Introduction}
Gravitational form factors (GFFs) provide valuable
information about the internal structure of hadrons, such as their spatial
distributions of mass~\cite{Ji:1994av,Kharzeev:2021qkd}, angular momentum~\cite{Ji:1996ek}, pressure, and 
shear forces~\cite{Polyakov:2002yz,Polyakov:2018zvc,Burkert:2018bqq,Shanahan:2018nnv,Kumericki:2019ddg,Lorce:2018egm,Burkert:2021ith,Meziani:2025dwu} -- for a review, see Ref.~\cite{Burkert:2023wzr}.
Recent measurements of such GFFs via deeply virtual Compton scattering
have generated a lot of theoretical and phenomenological activity.
To date, the GFFs of the proton have also been determined up to momentum
transfer ${-t} = 2$~GeV$^2$ from lattice QCD~\cite{Hackett:2023rif}.
The GFFs parameterize the graviton-proton vertex~\cite{Kobzarev:1962wt,Pagels:1966zza} and allow the construction
of the energy-momentum tensor (EMT) $T_{\mu\nu}$ of the proton~\cite{Polyakov:2002yz,Polyakov:2018zvc}
as the Wigner transform
of the QCD energy-momentum operator; the Wigner transform provides a map
from Hilbert space operators to quasi-probability distributions in
phase space.

In the context of semiclassical gravity, this proton EMT provides a
source for Einstein's equations of general relativity (GR) for the
metric of spacetime\footnote{{Of course, because Newton's constant $G
  \sim \ell_\mathrm{Planck}^2$, where $\ell_\mathrm{Planck}$ is the
  Planck length, modifications in spacetime sourced by a single proton
  are utterly negligible at distances probed at colliders. Our point
  here is purely formal, as we are not advocating that there are any
  relevant, observable gravitational effects sourced by the
  proton.}}. Alternatively, instead of reading Einstein's equations as
matter determining gravity ($g_{\mu\nu}$), one can read the equation
in reverse and determine how the spacetime with the desired properties
determines the EMT needed to produce it. This is the point of the
so-called \emph{energy conditions} (ECs), used to constrain the set of
possible EMTs that can source Einstein's equations
\cite{Hawking:1973uf}. Examples are the null energy condition (NEC),
the weak energy condition (WEC), the dominant energy condition (DEC),
and the strong energy condition (SEC)~\cite{Hawking:1973uf}. Though
not fundamental \cite{Visser:1995cc}, point-wise ECs have been
instrumental in general relativity over the years to rigorously
establish, e.g., incompleteness (``singularity") theorems and the area
law theorem for black holes \cite{Wald:1984rg}.  Violations of ECs are
widely known and, often, very consequential. SEC is a sufficient
condition to ensure the focusing of timelike or null geodesics so that
gravity is always attractive in GR \cite{Hawking:1973uf}, but this
condition is violated during inflation and the current accelerated
expansion of the universe. Other examples where point-wise ECs are
violated, usually because of quantum effects, can be found in
\cite{Roman:1986tp,Martin-Moruno:2017exc,Abdolrahimi:2016emo,Martin-Moruno:2021niw}. Nevertheless,
less stringent conditions obtained by imposing that ECs should only
hold on \emph{average}, such as the averaged null energy condition
(ANEC), have played an important role in general relativity
\cite{PhysRevD.17.2521,Roman:1986tp,PhysRevD.37.546,Borde:1987qr,Morris:1988tu,Friedman:1993ty,Graham:2007va,Witten:2019qhl}
and are known to hold in unitary, Lorentz-invariant, interacting
quantum field theories in Minkowski spacetime
\cite{Faulkner:2016mzt,Hartman:2016lgu}. {Thus, ECs
  provide an important tool for revealing the general physical
  properties of energy-momentum tensors \emph{even} when gravitational
  effects are negligible, as is the case here.}

The standard interpretation of the proton EMT as describing some medium, in the classical sense, with pressure and shear forces is only meaningful when the typical energy conditions hold\footnote{Standard Boltzmann kinetic theory satisfies all energy conditions because
the phase-space distribution function $f\geq 0$. Once quantum effects are considered by the replacement of the
classical phase-space distribution by the Wigner quasi-distribution,
which is not non-negative, energy condition violations can happen.
{ However, non-positivity of the Wigner distribution may occur even for type~I
EMTs and is unrelated to the Hawking-Ellis classification \cite{Hawking:1973uf}.}}. 
In fact, the mass radius of the proton \cite{Polyakov:2018zvc,Kharzeev:2021qkd}, $\langle r^2 \rangle_\mathrm{mass} = \int d^3 x \,\vec{x}^{\,2}\, T_{00}(\vec{x}\,)/m$ (where $m$ is the proton mass), only has straightforward interpretation when the local energy density is non-negative for all timelike observers, i.e., if WEC holds.  
It is therefore interesting to test the energy-momentum tensor
corresponding to the GFFs of the proton for potential violation of energy conditions. In this paper, we show that at large distances
from the center of the proton, all point-wise ECs appear to be satisfied, so the
dilute tails of the proton behave as ordinary matter. However, {using as input recent lattice results for the GFFs}, we find that near the center of the proton all point-wise ECs could be violated. A definitive answer
will require measurements or lattice QCD computations of GFFs to a somewhat
higher momentum transfer than is presently available. Finally, we show that ANEC leads to a new model-independent constraint for the proton {and pion} GFFs in QCD, which can be investigated on the lattice. 

{We remark that Jaffe has drawn attention to ambiguities in defining spatial charge densities 
in hadrons with a radius of order their Compton wavelength~\cite{Jaffe:2020ebz},
akin to issues due to
the identification of the Wigner transform of a local QFT operator
with a classical phase space distribution.
Also, Miller has discussed definitions of the electromagnetic ``radius" of the proton~\cite{Miller:2018ybm}. However, the points raised by these authors apply
the same to type~I EMTs such as for a spin-0 pion, and are
unrelated to the systematic and observer independent Hawking-Ellis classification \cite{Hawking:1973uf} we pursue. Thus, it is our goal here to extract from the EMT information about the causal eigenvectors and the corresponding mechanical interpretation. 
}

{ Finally, we would like to note that some of the implications of the point-wise (NEC, WEC, DEC, SEC)
energy conditions 
on model GFFs have  been considered previously in Ref.~\cite{Lorce:2018egm}, assuming an unpolarized proton where the spin vector $\vec\sigma_{ss}$ in the
expression for $T^{0i}$ in~(\ref{eq:EVs-etc}) is set to zero. 
However, the relevance of the Hawking-Ellis classification of the EMT
has not been realized before,
and Ref.~\cite{Lorce:2018egm} implicitly assumed the EMT to be of ordinary
type~I; we shall show below that for a proton with definite polarization the
EMT may turn out to be of type~IV in the core. 
Furthermore, as remarked above, in this work we 
derive novel model-independent constraints on the GFFs of spin 0 and spin 1/2 hadrons arising from ANEC, which have not been considered heretofore.
}
\\

\section{Gravitational form factors and the Wigner transform of the QCD energy-momentum operator}

The vertex of a $2^{++}$ glueball or graviton and a proton can
be parameterized in terms of three QCD
form factors~\cite{Kobzarev:1962wt,Pagels:1966zza,Polyakov:2002yz,Polyakov:2018zvc} as
\be \label{eq:Tmunu-FormFactors}
\left< p',s'\left| \, \hat T_{\mu\nu}(0)\, \right| p,s\right> =
\bar{u}_{s'}(p') \left[A(t)\frac{P_\mu P_\nu}{m} +
J(t)\frac{i(P_{\mu}\sigma_{\nu\rho} + P_{\nu}\sigma_{\mu\rho})\Delta^\rho}{2m}
+ D(t) \frac{\Delta_\mu\Delta_\nu-\eta_{\mu\nu}\Delta^2}{4m}
\right] u_s(p)~.
\ee
This represents the total energy-momentum tensor of quarks plus gluons which is
renormalization scale invariant; $\hat T_{\mu\nu}(0)$ denotes the Belinfante
symmetric QCD energy-momentum
operator at spacetime point 0.
Here, $P=(p'+p)/2$, $\Delta=p'-p$, and the Mandelstam variable $t=\Delta^2$;
in this section we shall be using the $\eta_{\mu\nu} = \mathrm{diag}(+,-,-,-)$ Minkowski metric in Cartesian coordinates
common in standard QFT textbooks~\cite{Peskin:1995ev}. Also, $s, s'$ denote the spin
states for a given polarization axis.
Note that $P\cdot\Delta=0$ because these four-momenta are on shell and, therefore, contracting the above matrix element with $\Delta^\mu$ gives zero, as this energy-momentum tensor is conserved.

The $A$ and $J$ FFs satisfy $A(0)=1$ so that the integral over all space of
the static EMT $T_{00}$ gives the mass of the proton, Eq.~(\ref{eq:intd3x_T00}) below.
Furthermore,
$J(0)=1/2$ is the spin of the proton. The $D$ FF is unconstrained except to satisfy
$tD(t)\to0$ as $t\to 0$.
For asymptotic momentum transfer, $-t \to \infty$, the form factors scale as follows~\cite{Tong:2022zax}:
\be
A(t) \sim \frac{1}{t^2}~,~~~ D(t) \sim \frac{1}{t^3}~,~~~ A(t) - 2J(t) \sim \frac{1}{t^3}~.
\label{asymptotics}
\ee
Note that the last expression implies a cancellation of the leading $\sim 1/t^2$ behavior between $A(t)$ and
$2J(t)$. Besides the asymptotic scaling laws, our current theoretical
knowledge of the FFs includes their determination from lattice QCD
up to momentum transfer ${-t} = 2$~GeV$^2$~\cite{Hackett:2023rif}, as already mentioned. 
Furthermore, they have also been computed 
in holographic models for the proton~\cite{Mamo:2019mka,Mamo:2021krl,Mamo:2022eui,Fujita:2022jus}.
For the electromagnetic FFs of the pion and kaon the transition to the asymptotic scaling
of perturbative QCD occurs at momentum transfer
of order several GeV~\cite{Ding:2024lfj}.

From the above matrix element, one can construct an
energy-momentum tensor on a fixed (but arbitrary) time
surface $x^0$~\cite{Polyakov:2002yz,Polyakov:2018zvc,Lorce:2018egm}:
\be \label{eq:T_munu(x)}
T_{\mu\nu}(\vec x\,) \,
{
= \frac{\left< x^0, \vec x \,\left|\, \hat T_{\mu\nu}(0) \,\right|\, x^0,\vec x \right>}
{\left< x^0, \vec x \,|\, x^0,\vec x \right>}
}
= \int\frac{\rmd^3\Delta}{2E\, (2\pi)^3}\,
e^{-i \vec \Delta\cdot\vec x}\,
\left<\frac{\vec\Delta}{2},s \left|\, \hat T_{\mu\nu}(0) \,\right| -\frac{\vec\Delta}{2},s\right>~,
\ee
with $E=\sqrt{\vec\Delta^2/4 + m^2}$ the energy of the on-shell proton states. Here,
the matrix element of $\hat T_{\mu\nu}(0)$ is taken for the same spin states in bra and ket.
The details of how to arrive at this expression are given in Appendix
\ref{app:semi-cl-em-tensor}.
In fact, $T_{\mu\nu}(\vec x\,)$ represents the Wigner transform (see also 
Ref.~\cite{Lorce:2018egm}) of the operator
$\hat T_{\mu\nu}(0)$ for momentum $\vec P =0$, so a more appropriate notation would be
$T_{\mu\nu}(\vec x, \vec P=0)$. The marginal of this quasi-probability distribution
corresponds to the expectation value
of $\hat T_{\mu\nu}(0)$ in a proton state with momentum $\vec P =0$ and spin $s$,
\be \label{eq:intd3x_T00}
\int\rmd^3x\,\, T_{\mu\nu}(\vec x\,) = \left< \vec P =0,s \left|
\hat T_{\mu\nu}(0) \right| \vec P =0,s \right> = m\, \delta_{\mu 0}\delta_{\nu 0}~.
\ee
Using the parameterization~(\ref{eq:Tmunu-FormFactors}) in~(\ref{eq:T_munu(x)}) we may express every component
of $T_{\mu\nu}(\vec x\,)$ in terms of the QCD form factors, see Refs.~\cite{Polyakov:2002yz,Polyakov:2018zvc,Lorce:2018egm} and appendix
\ref{sec:T_munu-A-J-D}.
\\

\section{Energy conditions} 
 
The point-wise energy conditions we will be considering are \cite{Wald:1984rg}:
\begin{itemize}
\item Null energy condition: $T_{\mu\nu}\ell^\mu \ell^\nu\geq 0$ for all null vectors $\ell^\mu$. 
\item Weak energy condition: $T_{\mu\nu}t^\mu t^\nu\geq 0$ for all timelike vectors $t^\mu$. The physical meaning here is that the energy density measured by any observer is non-negative. By continuity, this will also be true for any null vector, so WEC implies NEC. 
\item Dominant energy condition: for all future-directed timelike $t^\mu$, the vector $T^\mu_\nu t^\nu$ should be a future-directed non-spacelike vector. The physical meaning of this condition is that for any observer, the local energy density appears non-negative and the local energy flow vector is non-spacelike. One can prove that DEC implies that if $T_{\mu\nu}$ vanishes in some set region of spacetime, then it also vanishes on the future Cauchy development of that region \cite{Hawking:1973uf}. This is a necessary, but not sufficient, condition to ensure causality. Also, one can show that DEC implies WEC.
\item Strong energy condition: for all unit timelike $t^\mu$, $\left(T_{\mu\nu}- \frac{T}{2}\eta_{\mu\nu}\right)t^\mu t^\nu \geq 0$, where $T = \eta_{\mu\nu}T^{\mu\nu}$.
\end{itemize}

We employ the EMT of \eqref{eq:T_munu(x)} to write the energy conditions in terms of the following quantities:
\be
\begin{split}
T_{00} &= \int\frac{\rmd^3\Delta}{(2\pi)^3}\, e^{-i \vec \Delta\cdot\vec x}\,
\left[mA(t)-\frac{t}{4m}\left( A(t)-2J(t)+D(t)\right)\right]~,\\
M^i = T^{0i} &= -\frac{i}{2} \int\frac{\rmd^3\Delta}{(2\pi)^3}\,
e^{-i \vec \Delta\cdot\vec x}\, J(t) \, (\vec\Delta \times \vec\sigma_{ss})^i~,\\
P_t &= \frac{1}{2}\int\frac{\rmd^3\Delta}{(2\pi)^3}\, e^{-i \vec \Delta\cdot\vec x}\,
  \frac{t}{4m} D(t) \left(1+\left(\hat r \cdot \hat\Delta\right)^2\right)~,\\
P_r &= \int\frac{\rmd^3\Delta}{(2\pi)^3}\, e^{-i \vec \Delta\cdot\vec x}\,
  \frac{t}{4m} D(t) \left(1-\left(\hat r \cdot \hat\Delta\right)^2\right)~,\\
\Gamma &= \left(P_t +T_{00}\right)^2 - 4 \vec M^2~,\\
\rho &= \frac{1}{2}\left(T_{00} -P_t\right) + \frac{1}{2}\sqrt{\Gamma},\\
P_1 &= \frac{1}{2}\left(P_t -T_{00}\right) + \frac{1}{2}\sqrt{\Gamma}~.
\end{split}
\label{eq:EVs-etc}
\ee
In the second line, $\vec\sigma_{ss} = \xi^\dagger_s \vec\sigma \xi_s$, with $\xi_s$ the Pauli spinor of the proton.
For proton spin $+\frac{1}{2}$ along the $z$-axis we have $\xi_s=(1,0)^T$ and $\vec\sigma_{ss}=(0,0,1)=\hat z$. Also, our notation is $\vec{x} = r\,\hat r$.

The discriminant $\Gamma$
determines the type of energy-momentum tensor according to the Hawking-Ellis classification \cite{Hawking:1973uf}. 
In regions where $\Gamma >0$, the energy-momentum tensor is of type I \cite{Hawking:1973uf,Maeda:2018hqu}, one of the eigenvectors is timelike and the others are spacelike, and all eigenvalues are real. Physically, this means that the energy flux $T^{0i}$ can always be made to vanish via a local Lorentz transformation -- there is always a physical observer who finds no net energy flux in any direction -- so that the EMT is diagonal in a properly defined orthonormal frame \cite{Hawking:1973uf}. EMTs describing classical matter, where concepts such as pressure and shear stress have clear physical meaning, are expected to be of type I \cite{Hawking:1973uf,Wald:1984rg}. 
If $\Gamma = 0$, the tensor is of type II, two eigenvalues are degenerate corresponding to two null eigenvectors, and this case is typically associated with classical radiation \cite{Hawking:1973uf}. In general, in the absence of protecting symmetries, one expects that perturbations may take the system away from type II \cite{Martin-Moruno:2018eil}. Type III EMTs cannot occur in spherical symmetry \cite{Martin-Moruno:2021niw}, so this case is not relevant here. When $\Gamma <0$, the EMT is of type IV; one finds no causal eigenvectors, and two eigenvalues have a nonzero imaginary part. {
This can occur because in Lorentzian manifolds (such as Minkowski spacetime) the inner product between vectors can have any sign, which implies that the tensor used to determine the eigenvalues, $T_\mu^\nu$, does not need to have a real spectrum or be diagonalizable; complex conjugate pairs and Jordan blocks can appear \cite{Wald:1984rg}.}  Type IV cases are rather uncommon \cite{Roman:1986tp,Martin-Moruno:2013wfa,Abdolrahimi:2016emo}, and their properties away from the test-field limit in Einstein's equations were recently investigated in \cite{Maeda:2020dfp,Martin-Moruno:2021niw}.
For type IV EMTs, it is impossible for a physical observer not to experience a nonzero net energy flux. Furthermore, there are always local Lorentz frames in which the energy density is zero \cite{Hawking:1973uf}. Additionally, one can show that $T_{\mu\nu}t^\mu t^\nu$ cannot be bounded from below for all unit timelike vectors $t^\mu$ \cite{Roman:1986tp}. {Hence, type IV EMTs violate all the point-wise ECs mentioned above, since NEC is violated. This illustrates the importance of checking ECs, even in flat spacetime.} 

For completeness, we note that, away from the test-field case, it has been shown \cite{Maeda:2020dfp} that 
there is no hypersurface-orthogonal timelike Killing vector in a spacetimes sourced by type II, III, or IV EMTs. In particular, this implies that spacetimes generated by type IV EMTs cannot be static.  

Ref.\ \cite{Maeda:2022vld} gives necessary conditions that must be fulfilled to avoid violation of NEC, WEC, DEC, and SEC. In terms of the EMT quantities defined above, these energy conditions are necessarily violated in regions where
\be \label{eq:Gamma<0}
\Gamma < 0 ,\,\quad \mathrm{or} \qquad T_{00} + P_t <0~.
\ee
We will call these \emph{necessary conditions} because if satisfied, the energy conditions are violated. 
Eqs.~(\ref{eq:Gamma<0}) are important as they determine the type of EMT at hand, and the properties of
QCD matter and fields in the proton.

Now, let us focus on the case where $\Gamma>0$. Thus, the tensor is of type I, and the standard statements \cite{Wald:1984rg} for the energy conditions hold:
\begin{itemize}
\item NEC: $\rho+P_i \geq 0$ ($i=1,t,r$). This imposes the following constraints:
\be
\begin{split}
\rho+P_1 &= \sqrt{\Gamma}\geq 0~,\\
\rho + P_t &= \frac{1}{2}\left(T_{00} +P_t\right) + \frac{1}{2}\sqrt{\Gamma}\geq 0
~,\\
\rho+P_r &= \frac{1}{2}\left(T_{00} -P_t\right) + \frac{1}{2}\sqrt{\Gamma}+P_r \geq 0.
\end{split}
\ee
\item WEC: $\rho\geq 0$ and $\rho+P_i\geq 0$ ($i=1,t,r$), which is NEC. This implies that, in addition to the NEC conditions above, we also have
\begin{equation}
\frac{1}{2}\left(T_{00} -P_t\right) + \frac{1}{2}\sqrt{\Gamma}\geq 0.
\ee
\item DEC: $\rho\geq |P_i|$ ($i=1,t,r$). In other words, $\rho \geq 0$, $\rho +P_i\geq 0$, and $\rho -P_i\geq 0$ ($i=1,t,r$). This can be seen as WEC with the additional constraint $\rho - P_i\geq 0$. Here, this leads to
\be
\begin{split}
\frac{1}{2}\left(T_{00} -P_t\right) + \frac{1}{2}\sqrt{\Gamma} -P_t & \geq 0~,\\
\frac{1}{2}\left(T_{00} -P_t\right) + \frac{1}{2}\sqrt{\Gamma} -P_r  & \geq 0~,\\
T_{00} - P_t & \geq 0,
\end{split}
\ee
together with the conditions for WEC (we remind the reader that DEC implies WEC, which in turn implies NEC).
\item SEC: $\rho + \sum_{i=1}^3P_i \geq 0$ and $\rho + P_i \geq 0$ ($i=1,t,r$). This implies that
\be
P_t +P_r +\sqrt{\Gamma}  \geq 0,
\ee
must hold, in addition to the NEC conditions.
\end{itemize}

\section{Test of the energy conditions with QCD form factors}
We employ the parameterizations of
the form factors recently determined from state-of-the-art lattice QCD calculations~\cite{Hackett:2023rif}. These are dipole FFs such as
\be
A(t) = \frac{\alpha_A}{\left( 1 - t/\Lambda_A^2\right)^2}~,
\label{eq:FF-A(t)}
\ee
and similarly for $D(t)$ and $A(t)-2J(t)$.
The parameters $\alpha_i, \Lambda_i$ are given in the supplemental material of Ref.~\cite{Hackett:2023rif}. 
In that paper, it is shown that the
dipole forms hold at least up to momentum transfer $\Delta = \sqrt{-t} = \sqrt{2}$~GeV.
However, we do know that $D(t)$ as well as $A(t)-2J(t)$ asymptotically transition into a $\sim 1/t^3$ fall-off.
Hence, we restrict the dipole form for these two FFs to a maximum of $\Delta^*$; we then transition right away
to the asymptotic form with a coefficient chosen so that the FFs are continuous at $\Delta^*$ although the
first derivative will not be, of course. This is of no concern here since the energy conditions
involve integrals over the FFs.
We also know that $A(t)$ is supposed to fall off $\sim 1/t^2$ asymptotically so we assume that
(\ref{eq:FF-A(t)}) extends to asymptotic momentum transfer in that case.

\begin{figure}[htb]
  \begin{center}
  \includegraphics[scale = 0.65]{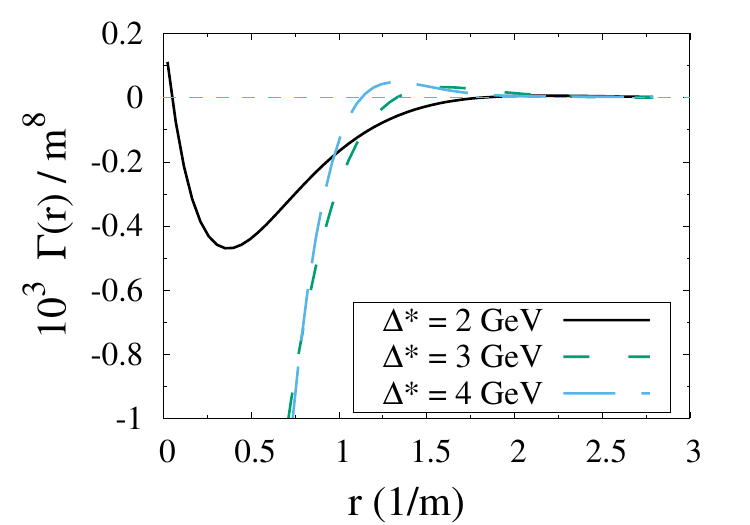}
  \includegraphics[scale = 0.65]{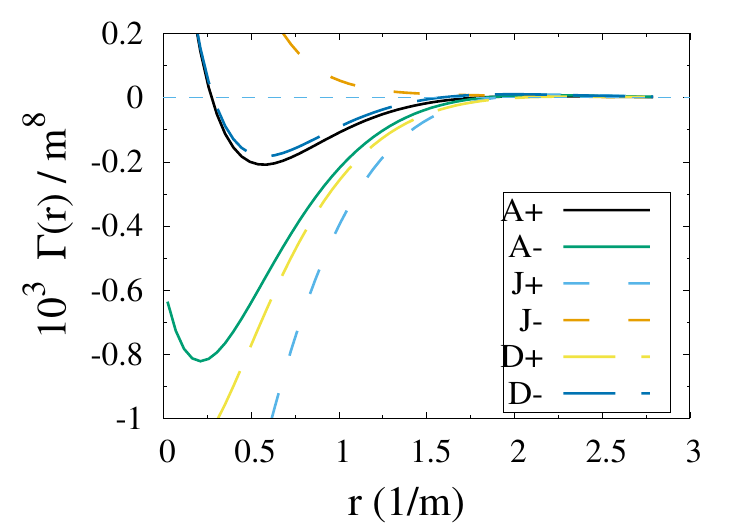}
  \end{center}
  \caption{Left: Plot of the discriminant $\Gamma(r)$ assuming lattice QCD FFs \cite{Hackett:2023rif} (with
  best fit parameters) for momentum
  transfer below $\Delta^*$, and a transition to the asymptotic FFs at $\Delta^*$. Here, $\vec x$ is chosen perpendicular
  to the direction of the spin axis, so $\vec M^2 > 0$. Radial distance is measured in units of the Compton
  wavelength $\bar\lambda=1/m$, where $m$ is the proton mass.
  Right: $\Gamma(r)$ for $\Delta^* = 2$~GeV obtained by variation of the pole masses in the individual GFFs within
  the uncertainties quoted in Ref.~\cite{Hackett:2023rif}.}
  \label{fig:Gamma}
\end{figure}
We first test the necessary conditions of Eq.~(\ref{eq:Gamma<0}). We find that  $T_{00} + P_t >0$ for all $r$
except possibly for $r$ very close to $0$ and large $\Delta^*$.
However, choosing $\vec x$ in the $x-y$ plane and the spin direction along $\hat z$, we find that the
discriminant $\Gamma(r)$ in the ``interior" region is mostly negative,
see Fig.~\ref{fig:Gamma} {(negative values are found whenever $\vec{x}$ is not perfectly aligned with the spin direction)}. This occurs because of the proton's $T^{0i}$ component, determined by the $J$ form factor. That is, in the proton's interior, the EMT is not of ``ordinary" type~I but of type~IV except in the ``tail" of the Wigner distribution of the proton. 
The farther out (in momentum transfer) the lattice QCD dipole FFs hold (greater $\Delta^*$), the greater $-\Gamma(r)$ becomes in the interior of the proton. 
\\

At radial distances sufficiently far from the center, we find $\Gamma(r) > 0$, so the tensor necessarily becomes type II before it turns into type~I.
For $\Delta^* = 2$~GeV this occurs at about two proton Compton wavelengths, dropping to $1.3\bar\lambda$ for
$\Delta^* = 3$~GeV and to $1.1\bar\lambda$ for $\Delta^* = 4$~GeV.
In the far tails of the Wigner distribution, all the energy conditions mentioned above are
satisfied. This can be seen from the expressions in Eq.~(\ref{eq:EVs-etc}): in the limit $r\to\infty$ we
obtain $T_{00}\sim m/r^3$ with a positive coefficient, while $M^i, P_t, P_r \to 0$ faster than $r^{-3}$; hence $\Gamma \to T_{00}^2$.

In the right panel of Fig.~\ref{fig:Gamma} we show various curves for $\Gamma(r)$. These were obtained by
variation of the pole masses $\Lambda_q$, $\Lambda_g$ in the individual GFFs within
the uncertainties quoted in Ref.~\cite{Hackett:2023rif}. For example, the curves labeled $A+$ ($A-$)
correspond to increasing (decreasing) both $\Lambda_{Aq}$ and $\Lambda_{Ag}$ simultaneously by the uncertainty on these
parameters quoted in the supplemental material of Ref.~\cite{Hackett:2023rif}. We do not vary
the $\alpha_{Aq}$, $\alpha_{Ag}$ parameters since $A(0)$, $J(0)$, and $D(0)$ correspond to conserved ``charges".
The figure shows that $\Gamma(r)>0$ throughout the core of the proton could be achieved only by the most
rapid allowed fall-off of $J(t)$ from $J(0)=1/2$. For $\Delta^* > 2$~GeV, a region of negative $\Gamma$
is unavoidable even in this case.
\\

\noindent
\section{New constraints on GFFs from ANEC} We now go beyond the strict point-wise EC considered above and investigate also the averaged null energy condition, which states that $\int d\zeta\, T_{\mu\nu}\ell^\mu \ell^\nu \geq 0$, where the integral is taken over a complete null geodesic with affine parameter $\zeta \in (-\infty,\infty)$ and tangent vector $\ell^\mu = dx^\mu/d\zeta$. In Minkowski spacetime, one can assume without any loss of generality $x^\mu(\zeta) = \zeta \ell^\mu$, with $\ell^\mu = (1,n^i)$, with $\vec{n}^2 = 1$. Using the full expression for the EMT (see Appendix \ref{sec:app-ANEC}), ANEC requires $\int\rmd\zeta \, \ell^\mu \ell^\nu \, T_{\mu\nu} = \int\rmd\zeta\, \left(
T^{00} - 2 n^i T^{0i} + n^i n^j T^{ij}
\right)\geq 0$. For $\vec{P}=0$, imposing that ANEC holds in QCD gives the following constraint on the $A$ and $J$ GFFs (see the derivation in Appendix \ref{sec:app-ANEC})
\be
\int_{-\infty}^0 dt \,
\left[ m A(t) - \frac{t}{4m}\left(A(t)-2 J(t)\right)
 \right] \geq 0.
 \label{newconstraint}
\ee
It is interesting to note that in some holographic models \cite{Mamo:2022eui} $A=2J$, so \eqref{newconstraint} holds
for $\int\rmd t\, A(t)\ge 0$ provided the mass $m\ge0$. 
In QCD, the integral in \eqref{newconstraint} is finite because $A - 2J \sim 1/t^3$ asymptotically \cite{Tong:2022zax}, but the $-t(A-2J)$ term is not positive definite~\cite{Hackett:2023rif}.   
Nevertheless, we have confirmed that this inequality indeed holds for the proton EMT when the model GFFs used in our calculations of the point-wise ECs are employed. 
Thus, ANEC implies\footnote{Besides the explicit proofs in Minkowski spacetime in \cite{Faulkner:2016mzt,Hartman:2016lgu}, it is known that violating ANEC would be generally catastrophic, as arguments exist stating that this could be used to build time machines \cite{Morris:1988tu,Friedman:1993ty} and violate the (generalized) second law of thermodynamics \cite{Wall:2009wi}.  Finally, we note that in Minkowski spacetime, all null lines are achronal so the ANEC condition discussed here is equivalent to the more stringent achronal averaged null energy condition \cite{Graham:2007va}, which is expected to hold even in curved spacetime.} that \eqref{newconstraint} provides a \emph{new model-independent constraint} that should be satisfied by proton GFFs in QCD.

{ We compare to ANEC for a spin-0 pion where the asymptotic $A(t)$ GFF takes
monopole form~\cite{Tanaka:2018wea,Tong:2021ctu}. We obtain
\be
\begin{split}
\frac{[\Delta^2 A(-\Delta^2)]_\infty}{4\pi \, |\vec{x}_{0\perp}|}
 +
\int\frac{\rmd^2\Delta_\perp}{(2\pi)^2}\, e^{-i \vec{\Delta}_\perp\cdot\vec{x}_{0\perp}}
\left[E A(-\Delta_\perp^2) - 
\frac{[\Delta^2 A(-\Delta^2)]_\infty}{2\Delta_\perp}\right] \ge 0
~.
\end{split}
\label{eq:ANEC_pi}
\ee
$\vec{x}_{0\perp}$ represents the displacement of the geodesic from the center
of the pion, and $[\Delta^2 A(-\Delta^2)]_\infty = \lim_{\Delta\to\infty}\, \Delta^2 A(-\Delta^2)$
with $\Delta = |\vec\Delta|$.
The first term in eq.~(\ref{eq:ANEC_pi}) is always positive and
ANEC$_\pi$ diverges as $\sim |\vec{x}_{0\perp}|^{-1}$ as $|\vec{x}_{0\perp}| \to 0$.
Hence, in this limit ANEC$_\pi \ge 0$ is satisfied.
However, for finite $|\vec{x}_{0\perp}|$, eq.~(\ref{eq:ANEC_pi}) represents
a new non-trivial constraint on the $A(t)$ GFF of the pion. For example,
if this GFF were described by a single monopole mass $\Lambda$ then
$m_\pi/\Lambda \gtrsim 0.21$ would be required by ANEC$_\pi \ge 0$.
}
\\

\section{Summary and Discussion}
The form factors of the graviton-proton vertex determine the Wigner transform of the
QCD energy-momentum tensor in the proton. Its spatial distribution has been used to investigate the mechanical properties of the proton, such as its pressure and shear forces. In this work, motivated by the fact that the EMT is a source term in Einstein's equations, we have checked its properties according to the classification of Hawking-Ellis using form factors 
determined from lattice QCD up to momentum transfer ${-t}=2$~GeV$^2$ where we transitioned to the
asymptotic forms expected from perturbative QCD. We have found that the EMT in the proton
is of ``ordinary" type~I in the radial tails beyond a few Compton wavelengths where the
standard energy conditions such as NEC, WEC, SEC, and DEC are satisfied. In the interior region, however, the proton EMT
 may be of type~IV, lacking any causal eigenvectors, and violating all the standard point-wise energy conditions. This unconventional behavior originates from the contribution of the $J$ form factor which determines the
$T^{0i}$ component of the EMT. In this regard, the EMT of the pion, where $T^{0i}=0$, will not be of type IV, though the pointwise ECs may still be violated in certain regions.

The exotic properties { potentially} displayed by the proton EMT, such as the fact that there may be regions within the proton where its energy density can vanish or become negative, call for some caution with definitions of the proton's mechanical properties, such as its mass radius\footnote{We refer to Ref.~\cite{Goharipour:2025yxm} for a recent comprehensive review
of various proton radius definitions.}, using this EMT. 
We propose using the point where the EMT transitions from type IV to type I (thus, necessarily becoming type II at that point) to define a new “gravitational radius” for the proton, which is free from the conceptual issues mentioned above. Using current input from the lattice, we obtain that this occurs at about $1-2$ Compton wavelength. The precise value of this ``classicalization" length scale above which typical mechanical properties within the proton are well-defined, standard ECs are fulfilled, and gravity is necessarily attractive, requires determining the gravitational form factors for momentum transfers beyond ${-t}=2$~GeV$^2$, either from lattice QCD or experiments. 

Finally, we showed that ANEC translates into the new model-independent, non-perturbative constraints  \eqref{newconstraint} { and \eqref{eq:ANEC_pi} for the proton $A(t)$
and $J(t)$ form factors and the $A(t)$ pion form factor in QCD, respectively.} A definitive confirmation of ANEC from data reaching to sufficiently high $-t$
could motivate additional theoretical work towards new rigorous proofs in asymptotically free theories\footnote{ANEC was proven to hold in unitary, Lorentz invariant quantum field theories in Minkowski spacetime in \cite{Faulkner:2016mzt} using quantum information-theoretic inequalities, such as the monotonicity of the relative entropy. In \cite{Hartman:2016lgu}, a proof was obtained using causality considerations.}.
In this context, it would be interesting to also
check other conditions, such as the quantum null energy condition, which has been thoroughly discussed in the literature \cite{Bousso:2015mna,Bousso:2015wca,Wall:2017blw,Balakrishnan:2017bjg}.

\begin{acknowledgments}
J.~N.\ thanks T.~Faulkner for discussions concerning the validity of average energy conditions in quantum field theories, and N.~Yunes for discussions about the meaning of the Hawking-Ellis energy-momentum tensor classification. A.~D.\ acknowledges support
by the DOE Office of Nuclear Physics through Grant DE-SC0002307 and the 
Illinois Center for Advanced Studies of the Universe (ICASU) at Urbana-Champaign for their
hospitality during a visit in March 2025, where this work was initiated. J.~N.\ is partly supported by the U.S. Department of Energy, Office of Science, Office for Nuclear Physics under Award No. DE-SC0023861.
\end{acknowledgments}

\appendix

\section{Wigner transform of the QCD energy-momentum operator}
\label{app:semi-cl-em-tensor}

We construct a proton state in position space as
\be \label{eq:|x>}
|x\rangle = \int\limits_{p} e^{-ip\cdot x}\, |p\rangle~.
\ee
Here, $|p\rangle$ refers to a proton state in momentum space, i.e.\ an eigenstate
of the momentum operator. These
are on-shell states with $p^0=\sqrt{\vec p^2 + m^2}$ implicitly understood;
hence, the integration is only over three-momentum $\vec p$ with
integration measure~\cite{Peskin:1995ev}
\be
\int\limits_{p} = \int \frac{\rmd^3p}{\sqrt{2p^0} \, (2\pi)^3}~.
\ee
We may then compute $\langle y|x\rangle$
using the covariant momentum space normalization
\be
\langle p' | p \rangle = 2p^0\, (2\pi)^3 \, \delta(\vec p\,^\prime - \vec p\,)~,
\ee
As these are on-shell states, the $\delta$-function for three-momentum
in effect also enforces $p^0 = p^{\prime 0}$.
With this,
\be \begin{split}
\langle y|x\rangle &= \int\limits_{p} \int\limits_{p'}
e^{-ip\cdot x}\, e^{ip'\cdot y}\, \langle p' | p \rangle = \int \frac{\rmd^3p}{(2\pi)^3} \, e^{-ip^0 (x^0 - y^0)}\,
  e^{i\vec p\cdot (\vec x - \vec y)}
\end{split}
\ee
On a surface of fixed time, $y^0=x^0$,
\be
\langle x^0,\vec y\, |\, x^0,\vec x\rangle = \delta(\vec x - \vec y)~.
\ee
\\

Let us now consider the equal time matrix element
\be
\left< x^0, \vec x \left|\, \hat T_{\mu\nu}(0) \, \right| x^0, \vec x\right>~.
\ee
Using (\ref{eq:|x>}), this becomes
\be
\begin{split}
\int_p\int_{p'} e^{-i p\cdot x} e^{i p'\cdot x} \left<p'\left|\, \hat T_{\mu\nu}(0) \,\right| p\right>  &=
\int\rmd^3P \int\frac{\rmd^3\Delta}{\sqrt{2p^0}\, (2\pi)^3 \sqrt{2p^{\prime 0}}\, (2\pi)^3}\,
e^{i \Delta^0 x^0} e^{i (\vec P - \frac{1}{2}\vec \Delta)\cdot\vec x}\,
e^{-i (\vec P + \frac{1}{2}\vec \Delta)\cdot\vec x}\,
\left<p' \left|\, \hat T_{\mu\nu}(0) \, \right| p\right>~.
\end{split}  \label{eq:matrix-element-1}
\ee
Here we introduced the four-vectors
\be
P = \frac{p+p'}{2}~,~~~  \Delta=p'-p~.
\ee
Due to the change of integration variables,
in~(\ref{eq:matrix-element-1}) we should now read $p^0$ and $p^{\prime 0}$ as shorthands for
\be \begin{split}
p^0 &= \sqrt{(\vec P-\vec\Delta/2)^2 + m^2} \equiv E~,~~~~
p^{\prime 0}= \sqrt{(\vec P+\vec\Delta/2)^2 + m^2} \equiv E'~,
\end{split}  \label{eq:kinematics}
\ee
with $\Delta_0 = E'-E$ and $P_0 = (E'+E)/2$. Now we use that
\be
\int\frac{\rmd^3P}{(2\pi)^3} \, e^{i P \cdot(x-x)} = \left<x^0,\vec x\left. \right| x^0, \vec x\right>~,
\ee
and interpret $\left<x^0,\vec x\left. \right| x^0, \vec x\right>$ in the sense of distributions to see that its inverse may be represented as $(2\pi)^3 \delta(\vec{P}-\vec{Q})$
when working with momentum space basis vectors. Thus, we find that 
\be
\left< x^0, \vec x \left|\, \hat T_{\mu\nu}(0) \, \right| x^0, \vec x\right>
= \left<x^0,\vec x\left. \right| x^0, \vec x\right>
\int\frac{\rmd^3\Delta}{(2\pi)^3}\,
\frac{1}{\sqrt{4 P_0^2 - \Delta_0^2}} \, e^{i\Delta \cdot x}\,
\left<\vec P + \frac{\vec\Delta}{2},s \left|\, \hat T_{\mu\nu}(0) \,\right| \vec P -\frac{\vec\Delta}{2},s\right>~,
\ee

The final step is to identify the expectation value of the energy-momentum tensor as
\be \label{eq:T_munu(x0,x)_Appendix}
T_{\mu\nu}(x^0,\vec x, \vec P) = \frac{\left< x^0, \vec x \left|\, \hat T_{\mu\nu}(0) \, \right| x^0, \vec x\right>}
{\left<x^0,\vec x\left. \right| x^0, \vec x\right>} =
\int\frac{\rmd^3\Delta}{(2\pi)^3}\,\frac{1}{\sqrt{4E' E}} \,
e^{i (E'-E) x^0}\, e^{-i\vec\Delta \cdot \vec x}\,
\left<\vec P + \frac{\vec\Delta}{2},s \left|\, \hat T_{\mu\nu}(0) \,\right| \vec P -\frac{\vec\Delta}{2},s\right>~.
\ee
The static EMT of Polyakov {\it et al.} is recovered by taking $\vec P
=0$, the ``Breit frame", which gives the energy-momentum tensor in
Eq.~(\ref{eq:T_munu(x)}) of the main text:
\be
T_{\mu\nu}(\vec x\,)
= \int\frac{\rmd^3\Delta}{2E\, (2\pi)^3}\,
e^{-i \vec \Delta\cdot\vec x}\,
\left<\frac{\vec\Delta}{2},s \left|\, \hat T_{\mu\nu}(0) \,\right| -\frac{\vec\Delta}{2},s\right>.
\ee
Eq.~(\ref{eq:T_munu(x0,x)_Appendix}) represents the Wigner-Weyl transform of $\hat T_{\mu\nu}(0)$
to the phase-space quasi-probability distribution $T_{\mu\nu}(x^0,\vec x,\vec P)$.
Note that this is conserved,
\be
\partial^\mu T_{\mu\nu}(x^0,\vec x,\vec P) =
\int\frac{\rmd^3\Delta}{(2\pi)^3}\,\frac{1}{\sqrt{4E' E}} \,
e^{i (E'-E) x^0}\, e^{-i\vec\Delta \cdot \vec x}\,
i\Delta^\mu
\left<\vec P + \frac{\vec\Delta}{2},s \left|\, \hat T_{\mu\nu}(0) \,\right| \vec P -\frac{\vec\Delta}{2},s\right>
=0 ~,
\ee
where $\Delta^\mu = (\Delta^0,\vec\Delta)$.

Integrating (\ref{eq:T_munu(x0,x)_Appendix}) over all space,
\be
T_{\mu\nu}(\vec P) =
\int\rmd^3 x\, T_{\mu\nu}(x^0,\vec x,\vec P) = \frac{1}{2E}
\left<\vec P,s \left|\, \hat T_{\mu\nu}(0) \,\right| \vec P,s\right>
 = \frac{P_\mu P_\nu}{m}~.
\ee
On the r.h.s.\ both $E$ and $P_0$ are evaluated for $\vec\Delta=0$, so that $P_0=E$.
The EMT for an ensemble of such particles characterized in infinite volume by the classical
momentum-space mass density distribution $mf(\vec P) \ge 0$ is
\be
T_{\mu\nu}^\mathrm{(KT)} = \int \frac{\rmd^3 P}{2E}\, T_{\mu\nu}(\vec P) \, m f(\vec P)=
\int \frac{\rmd^3 P}{2E}\, P_\mu P_\nu\, f(\vec P)~.
\ee
This is the EMT of classical kinetic theory.

\section{Components of the Wigner EMT in terms of the form factors}
\label{sec:T_munu-A-J-D}

We now write down the components of the static, Breit frame EMT
$T_{\mu\nu}(\vec x)$ in terms of the form factors. The matrix
element of $\hat T_{00}(0)$ is
\be \label{eq:hat-T_00}
\left<\frac{\vec\Delta}{2},s \left|\, \hat T_{00}(0) \,\right| -\frac{\vec\Delta}{2},s\right> =
2mE\left[A(t)+\frac{-t}{4m^2}\left( A(t)-2J(t)+D(t)\right)\right]~.
\ee
The terms $\sim A(t), D(t)$ can be reproduced easily using $\bar u_{s'}(p') u_s(p) = 2E
\delta_{s' s} = 2\sqrt{\vec\Delta^2/4 + m^2}\delta_{s' s}$, for
$\vec p^{\, \prime}+\vec p = 0$ and $\vec p = -\vec\Delta/2$. 
The simplest approach for the term $\sim J(t)$ is to use the Gordon identity in the form
\be
\bar{u}_s(p') \frac{i\sigma^{0\nu}\Delta_\nu}{2m} u_s(p) = \bar{u}_s(p') \gamma^0 u_s(p) - 
\frac{P^0}{m}  \bar{u}_s(p') u_s(p)~.
\ee
The second term on the r.h.s.\ immediately gives $-2E^2/m$. For the first term on the r.h.s., using the
chiral/Weyl representation of $\gamma^0$~\cite{Peskin:1995ev}, we have $\gamma^0 u_s(p) = u_s(p')$ since
$\vec p = - \vec p^{\, \prime}$; so, for that term one obtains $\bar{u}_s(p') u_s(p')=2m$. In all, the r.h.s.\
evaluates to $-2\vec{p}^{\,2}/m = -(1/2) \vec \Delta^2/m = t/2m$. Hence,
\be
\bar{u}_s(p') ~
J(t)\frac{2iP_0 \, \sigma_{0\rho} \Delta^\rho}{2m}~ u_s(p) = 2J\, \frac{tE}{2m}~,
\ee
which agrees with the term $\sim J(t)$ in Eq.~(\ref{eq:hat-T_00}).

From Eq.~(\ref{eq:T_munu(x)}) then,
\be \begin{split}
T_{00}(r) &= m \int\frac{\rmd^3\Delta}{(2\pi)^3}\, e^{-i \vec \Delta\cdot\vec x}\,
\left[A(t)+\frac{-t}{4m^2}\left( A(t)-2J(t)+D(t)\right)\right] ~.
\label{defineT_00}
\end{split}
\ee
Here $r=|\vec x|$ because the integral does not depend on the direction of $\vec x$.
Integrating over all of space confirms that
\be
\int \rmd^3 x \, T_{00}(r) = m~,
\ee
since $A(0)=1$. 
\\

We move on to $T^{0i}(\vec x)$. The matrix element of $\hat T^{0i}(0)$ is given by Eq.~(17c) of
Ref.~\cite{Polyakov:2018zvc}:
\be
\left< p',s\left| \, \hat T^{0i}(0)\, \right| p,s\right> =
- E \, J(t) \,i\, (\vec\Delta \times \vec\sigma_{ss})^i~.
\ee
To confirm, we provide a few details of the derivation. We again use the Gordon identity in
the form
\be
\bar{u}_s(p') \frac{i\sigma^{i\nu}\Delta_\nu}{2m} u_s(p) = \bar{u}_s(p') \gamma^i u_s(p)
= - {u^\dagger}_s(p') \gamma^i u_s(p')~.
\ee
Here we used that $\gamma^0$ and $\gamma^i$ anti-commute, and $\gamma^0 u_s(p) = u_s(p')$.
To proceed, we employ the following representation of Dirac matrices and spinor solutions~\cite{Peskin:1995ev}:
\be
\gamma^i = \begin{pmatrix}
0 & \sigma^i \\
-\sigma^i & 0
\end{pmatrix}  ~,~
u_s(p) = \begin{pmatrix}
    \sqrt{p\cdot\sigma} \, \xi_s \\
    \sqrt{p\cdot\bar\sigma} \, \xi_s
\end{pmatrix}  ~,~
u^\dagger_s(p) = \begin{pmatrix}
    \xi_s^\dagger\, \sqrt{p\cdot\sigma} \\
    \xi_s^\dagger \, \sqrt{p\cdot\bar\sigma}
\end{pmatrix}
~.
\ee
Here, $\sigma^\mu = (1,\vec\sigma)$ and $\bar\sigma^\mu = (1,-\vec\sigma)$.
The Pauli spinors
\be
\xi_{\uparrow} = 
\begin{pmatrix}
\cos\frac{\theta}{2} \\
e^{i\phi}\sin\frac{\theta}{2}
\end{pmatrix}~, ~~~
\xi_{\downarrow} = 
\begin{pmatrix}
-e^{-i\phi}\sin\frac{\theta}{2}\\
\cos\frac{\theta}{2}
\end{pmatrix}
\ee
describe the spin component along the axis $\hat n = (\sin\theta \cos\phi, \sin\theta\sin\phi,\cos\theta)$.
We now use the Pauli matrix identities
\be
\sqrt{p\cdot\sigma} = \sqrt{E} \left( 
\frac{w_- + w_+}{2} + \hat{p}\cdot\vec\sigma \frac{w_- - w_+}{2}\right)~,~~
\sqrt{p\cdot\bar\sigma} = \sqrt{E} \left( 
\frac{w_- + w_+}{2} + \hat{p}\cdot\vec\sigma \frac{w_+ - w_-}{2}\right)~,~~~
w_\pm = \sqrt{1\pm |\vec p|/E}
\ee
to evaluate
\be
u^\dagger_s(p') \gamma^i u_s(p') = p^{\prime\,\ell} \xi_s^\dagger \left[ 
\sigma^i, \sigma^\ell \right]\, \xi_s \equiv i \left(\vec \Delta \times \vec\sigma_{ss}\right)^i~.
\ee
Here, $\sigma_{ss}^i = \xi^\dagger_s \sigma^i \xi_s$.
For proton spin $+\frac{1}{2}$ along the $z$-axis (i.e.\ $\theta=0$ so that $\hat n = \hat z$, and
$\xi_s = \xi_\uparrow$) we have $\vec\sigma_{ss}=(0,0,1)$ and
$(\vec\Delta \times \vec\sigma_{ss})^x=\Delta^y$, $(\vec\Delta \times \vec\sigma_{ss})^y=-\Delta^x$,
$(\vec\Delta \times \vec\sigma_{ss})^z=0$.
Then, from Eq.~(\ref{eq:T_munu(x)}),
\be
T^{0i}(\vec x) = -\frac{i}{2} \int\frac{\rmd^3\Delta}{(2\pi)^3}\,
e^{-i \vec \Delta\cdot\vec x}\, J(t) \, (\vec\Delta \times \vec\sigma_{ss})^i ~.
\label{eq:T_0i}
\ee
Specifically for spin $+\frac{1}{2}$ along the $z$-axis,
\be
\begin{split}
T^{0y}(x,y,z) = - T^{0x}(y,x,z) &= \frac{i}{2} \int\frac{\rmd^3\Delta}{(2\pi)^3}\,
e^{-i \vec \Delta\cdot\vec x}\, \Delta^x\, J(t)~,~~~~
T^{0z} = 0~.
\end{split}
\ee
\\

Finally, we write the components of $T^{ij}(\vec x)$, using $\vec P=0$ and $\bar u_s(p') u_s(p) = 2E$:
\be
T^{ij}(\vec x) = \int\frac{\rmd^3\Delta}{(2\pi)^3}\, e^{-i \vec \Delta\cdot\vec x}\, D(t)\,
\frac{\Delta^i \Delta^j - \delta^{ij}\vec\Delta^2}{4m}~.
\ee
%

\section{Derivation of the eigenvalues of $T_{\mu\nu}$}
\label{sec:T_munu-EVs}

In this appendix, we adhere to the common mostly plus choice for the metric tensor $g_{\mu\nu}$ used in the GR literature \cite{Wald:1984rg}. The Minkowski metric in Cartesian coordinates is now $\eta_{\mu\nu}  = \mathrm{diag}(-1,1,1,1)$. 

In semiclassical general relativity, in principle, any smooth Lorentzian manifold can be a spacetime for a suitably determined source given by the energy-momentum tensor $T_{\mu\nu}$, such that Einstein's equations
\be
R_{\mu\nu} - \frac{1}{2}g_{\mu\nu}R = 8 \pi G\, T_{\mu\nu}
\label{Einsteins}
\ee
are fulfilled. Above,  $T_{\mu\nu}(x) = \langle \hat{T}_{\mu\nu}(x) \rangle $ is the expectation value of the energy-momentum tensor operator in a quantum state (with pure vacuum contributions subtracted, so we are only interested in how ``matter" curves spacetime). Gravity can be treated classically away from large curvature regions, such as near black hole singularities, where intrinsic quantum effects in gravity cannot be neglected. 

Clearly, instead of reading \eqref{Einsteins} as matter ($T_{\mu\nu}$) determines gravity ($g_{\mu\nu}$), one can read the equation in reverse and determine how the spacetime with its desired properties (e.g., absence of closed timelike curves) determines the $T_{\mu\nu}$ needed to produce it. This is the idea behind the so-called energy conditions \cite{Martin-Moruno:2017exc}, which may be used to constrain the set of possible energy-momentum tensors that appear in Einstein's equations \cite{Hawking:1973uf}. 

As mentioned in the main text, the energy conditions we consider are the null energy condition, the weak energy condition, the dominant energy condition, and the strong energy condition, whose definitions were given in the main text, and are also be found in \cite{Wald:1984rg}.

Let us now show how these energy conditions are translated into conditions for the eigenvalues of the energy-momentum tensor. We first note that $T^\mu_\nu$ need not be symmetric. Furthermore, even though $T_{\mu\nu}$ is a symmetric real (0,2)-tensor, since $g_{\mu\nu}$ is not positive-definite, the linear map $T^{\mu}_\nu$ from vectors into vectors is not guaranteed to be diagonalizable in the sense that it need not have four linearly independent eigenvectors. This is discussed in detail in Hawking and Ellis' classic book \cite{Hawking:1973uf}. Of particular relevance for the case of the proton is the so-called type I energy-momentum tensors, where one of the eigenvectors of $T^\mu_\nu$ is timelike, with the others being spacelike. This is the case of energy-momentum tensors describing ordinary matter, such as a perfect fluid. The other possibilities are: type II, in which $T^\mu_\nu$ has a double null eigenvector; type III, where one finds a triple null eigenvector; and type IV, in which $T^\mu_\nu$ has no timelike or null eigenvector (in fact, it has eigenvalues with nonzero imaginary parts). The standard interpretation of mechanical quantities within the proton, such as its pressure and shear forces, inherently assumes that the energy-momentum tensor in the proton is of type I.  

The energy-momentum tensor $T^{\mu\nu}$ of the proton, computed in the previous section, has the following nonzero components in Cartesian coordinates
\be
T^{00}(r)  = m \int\frac{\rmd^3\Delta}{(2\pi)^3}\, e^{-i \vec \Delta\cdot\vec x}\,
\left[A(t)+\frac{-t}{4m^2}\left( A(t)-2J(t)+D(t)\right)\right],
\ee
\be
T^{0i}(\vec{x}) = -\frac{i}{2} \int\frac{\rmd^3\Delta}{(2\pi)^3}\,
e^{-i \vec \Delta\cdot\vec x}\, J(t) \, (\vec\Delta \times \vec\sigma_{ss})^i,
\ee
\be
T^{ij}(r) = \int\frac{\rmd^3\Delta}{(2\pi)^3}\, e^{-i \vec \Delta\cdot\vec x}\, D(t)\,
\frac{\Delta^i \Delta^j - \delta^{ij}\vec\Delta^2}{4m}~=\left(\frac{x^i x^j}{r^2}-\frac{\delta^{ij}}{3}\right)s(r) + \delta^{ij}p(r),
\ee
where 
\be
\begin{split}
s(r) &= -\frac{1}{4m}r\frac{d}{dr}\left(\frac{1}{r}\frac{d}{dr}\mathcal{D}(r)\right) =
\frac{3}{2}
\int\frac{\rmd^3\Delta}{(2\pi)^3}\, e^{-i \vec \Delta\cdot\vec r}\, D(t) 
\frac{(\hat r\cdot\vec\Delta)^2-\frac{1}{3}\vec\Delta^2}{4m}~,\\
p(r) &= \frac{1}{6m}\frac{1}{r^2}\frac{d}{dr}\left(r^2\frac{d}{dr}\mathcal{D}(r)\right) = 
\frac{1}{6m}\nabla^2 \mathcal{D}(r) =
\frac{1}{6m} \int\frac{\rmd^3\Delta}{(2\pi)^3}\, e^{-i \vec \Delta\cdot\vec r}\, tD(t)~
,
\end{split}
\ee
and $\mathcal{D}(r)$ represents the Fourier transform of $D(t)$.

To see how the energy conditions dictate the behavior of the eigenvalues of $T^\mu_\nu$, we use the fact that any Lorentzian manifold is locally flat to introduce a vierbein (or tetrad) $\{e_{a}^\mu(x)\}$, with $a=0,1,2,3$, which define an orthonormal basis such that 
\be
g_{\mu\nu}(x)e_{a}^\mu(x) e_{b}^\nu(x) = \eta_{ab}
\label{defineorthonormal}
\ee
and
\be
g_{\mu\nu}(x) = \eta_{ab}\,e^{a}_\mu(x) e^{b}_\nu(x),
\ee
where $\eta_{ab} = \mathrm{diag}(-1,1,1,1)$.
Note that we raise and lower $\mu$ and $\nu$ indices using $\eta_{\mu\nu}$, and $a$ and $b$ indices using $\eta_{ab}$. One can now determine any tensor in the local Minkowskian frame. For example, for the energy-momentum tensor, we define
\be
T_{ab} = T_{\mu\nu}\,e^\mu_{(a)} e^\nu_{(b)}.
\ee
We use spherical coordinates so $g_{\mu\nu} = \mathrm{diag}(-1,r^2 \sin^2 \theta,r^2,1)$ and
\be
e^\mu_{(0)} = \begin{pmatrix}
  1 \\
  0 \\
  0\\
  0
  \end{pmatrix},\qquad e^\mu_{(1)} = \begin{pmatrix}
  0 \\
  \frac{1}{r \sin\theta} \\
  0\\
  0
  \end{pmatrix},\qquad e^\mu_{(2)} = \begin{pmatrix}
  0 \\
  0 \\
  \frac{1}{r}\\
  0
  \end{pmatrix},\qquad e^\mu_{(3)} = \begin{pmatrix}
  0 \\
  0 \\
  0\\
  1
  \end{pmatrix}.
\ee
Note that \eqref{defineorthonormal} is satisfied. 
In spherical coordinates, the spatial part of the energy-momentum tensor is diagonal. In fact, the only nonzero components are $T^{00}$ and 
\be
\begin{split}
T^{\phi\phi} = \frac{1}{r^2 \sin^2 \theta}\left(p(r) - \frac{s(r)}{3}\right) \equiv \frac{1}{r^2 \sin^2 \theta} P_t
~ &,~~~
T^{\theta\theta} = \frac{1}{r^2} P_t\\
T^{rr} \equiv P_r = p(r) + \frac{2}{3}s(r)~ &, ~~~
T^{0\phi} = -\frac{1}{2r} \, \frac{d}{dr}\mathcal{J}(r)~.
\end{split}
\ee
Here, we took the spin of the proton in the $z$ direction.

Even before computing the eigenvalues of the energy-momentum tensor, one can see that this tensor cannot be of type III \cite{Hawking:1973uf}, though it can in principle be of types I, II, or IV. One can now find the eigenvalues $\lambda$ using this spherical coordinate system by solving the following equation
\be
\det \left(T^{a}_b-\lambda\, \eta^{a}_{b}\right)=0.
\ee
Besides the obvious $P_r$ and $P_t$, the other two eigenvalues are
\be
\lambda_\pm = \frac{1}{2}\left(P_t -T_{00}\right) \pm \frac{1}{2}\sqrt{\left(P_t +T_{00}\right)^2 - \sin^2\theta \left(\frac{d}{dr}\mathcal{J}(r)\right)^2} = \frac{1}{2}\left(P_t -T_{00}\right) \pm \frac{1}{2}\sqrt{\left(P_t +T_{00}\right)^2 - 4 (T^{0i})^2}.
\ee
The discriminant
\be
\Gamma = \left(P_t +T_{00}\right)^2 - 4 (T^{0i})^2
\label{discriminant}
\ee
determines the properties of the energy-momentum tensor. When $\Gamma >0$, the energy-momentum tensor is of type I \cite{Hawking:1973uf,Maeda:2018hqu}. If $\Gamma = 0$, the tensor is of type II. When $\Gamma <0$, the energy-momentum tensor is of type IV. 

Denoting the timelike eigenvector by $t^\mu$ and its eigenvalue by $\rho$, note that 
\be
T^\mu_\nu t^\nu= -\rho\, t^\mu, 
\ee
because of our metric signature. Thus,  $\rho = -\lambda_{-}$ is the rest energy density \cite{Maeda:2022vld}. Therefore, the quantities we use to determine the energy conditions in the main text are
\be
\begin{split}
\rho &= \frac{1}{2}\left(T_{00} -P_t\right) + \frac{1}{2}\sqrt{\Gamma}~,~~~
P_1 = \frac{1}{2}\left(P_t -T_{00}\right) + \frac{1}{2}\sqrt{\Gamma}~,\\
P_2 &= P_t~,~~~
P_3 = P_r~.
\end{split}
\ee

\section{Details about the ANEC calculation}
\label{sec:app-ANEC}

We first consider the averaged null energy condition for the
spin-$1/2$ proton and then for the spin-0 pion.

The ANEC integral for the null geodesic $\ell^\mu=(1,n^i)$ is
\be
\int\limits_{-\infty}^\infty\rmd\zeta \, \ell^\mu \ell^\nu \, T_{\mu\nu} = 
\int\limits_{-\infty}^\infty\rmd\zeta\, \left(
T^{00} - 2 {n}^i T^{0i} + {n}^i {n}^j T^{ij}
\right)~.
\ee
Using the general expression in \eqref{eq:T_munu(x0,x)_Appendix}, this can be expressed as
\be
\begin{split}
\int\frac{\rmd^3\Delta}{(2\pi)^2} \,
\frac{\delta\left(\Delta_\mu \ell^\mu\right)}{\sqrt{4EE'}}
\left[ 
\bar{u}(p') u(p) \, \frac{A(t)}{m} \left(\ell\cdot P\right)^2  +
\frac{J(t)}{m}\, \ell_\mu P^\mu\,  \ell_\nu \Delta_\rho \, \bar{u}(p') \, i\sigma^{\nu\rho} u(p) 
\right].
\end{split}
\ee
We note that the $D$ form factor does not contribute to this integral.
The spinor matrix elements for general $\vec P$ have
been computed in \cite{Lorce:2017isp,Lorce:2018egm},
\be
\begin{split}
\bar{u}(p') u(p) &= {\cal N}^{-1}\left[
2\left(P_0^2 - \vec{P}^2 + P_0m\right) + i \epsilon^{0ijk}P_i \Delta_j S_k
\right] = {\cal N}^{-1}\left[
2\left(P_0^2 - \vec{P}^2 + P_0m\right) - i \vec P \cdot (\vec\Delta \times \vec S)
\right]~,\\
\bar{u}(p') i\sigma^{\mu\rho}\Delta_\rho u(p) &= 
{\cal N}^{-1}\left\{
P^\mu\Delta^2 + m(\eta^{\mu 0}\Delta^2-\Delta^\mu\Delta^0) 
+ 2\left[(P^0+m)i\epsilon^{\mu\nu\rho\lambda}\Delta_\nu P_\rho S_\lambda -
\frac{\Delta^2}{4} i\epsilon^{\mu\nu\rho0}\Delta_\nu S_\rho 
- (P\cdot S) i\epsilon^{\mu\nu\rho0}\Delta_\nu P_\rho
\right]
\right\}~
\end{split}
\ee
with $S^\mu=(0,\vec S)$ the rest frame spin vector; 
in the notation of Refs.~\cite{Polyakov:2002yz,Polyakov:2018zvc} $\vec S$ had previously been denoted $\vec\sigma_{ss}$,
up to sign.
Furthermore, ${\cal N} = \sqrt{p^0+m}\sqrt{p^{\prime 0}+m} =
\sqrt{E+m}\sqrt{E'+m}$ and $P^0=(p^0+p^{\prime 0})/2 = (E+E')/2$.
We can also write the latter for $\mu=0$ and $\mu=i$ as
\be
\begin{split}
\bar{u}(p') i\sigma^{0\rho}\Delta_\rho u(p) &= {\cal N}^{-1}\left\{
P^0\Delta^2 - m\vec\Delta^2 + 2(P^0+m) i \vec{P}\cdot(\vec\Delta\times\vec S)\right\}~,\\
\bar{u}(p') i\sigma^{i\rho}\Delta_\rho u(p) &= {\cal N}^{-1}\left\{
P^i\Delta^2 -m\Delta^i\Delta^0 
+2\left[i(P^0+m)\left(P^0(\vec\Delta\times\vec S)^i-\Delta^0(\vec P\times\vec S)^i\right)
+\frac{\Delta^2}{4}i (\vec\Delta\times\vec S)^i -
(\vec{P}\cdot\vec{S}) i (\vec\Delta\times\vec P)^i
\right]\right\}~.
\end{split}
\label{eq:ubar-sigma-u_components}
\ee
The ANEC integral then becomes
\be
\begin{split}
\int\frac{\rmd^3\Delta}{(2\pi)^2} \,
\frac{\delta\left(\Delta_\mu \ell^\mu\right)}{\sqrt{4EE'}}
\left[ 
\bar{u}(p') u(p) \, \frac{A(t)}{m} \left(\ell\cdot P\right)^2  +
\frac{J(t)}{m}\, \ell_\mu P^\mu\,  \ell_\nu \Delta_\rho \, \bar{u}(p') \, i\sigma^{\nu\rho} u(p) 
\right].
\end{split}
\ee
The integrand can be simplified by considering the transformation
$\vec\Delta \to - \vec\Delta$. This lets $\Delta^0 \to - \Delta^0$
since $E \leftrightarrow E'$, and $P^0 \to P^0$. Therefore, the term
in $\bar{u}(p') u(p)$ involving $\vec\Delta \times \vec S$ is odd and
drops out. Along the same lines, all terms in $\Delta_\rho \,
\bar{u}(p') \, i\sigma^{\nu\rho} u(p)$ which are odd under $\vec\Delta
\to - \vec\Delta$ can be dropped. The ANEC integral then simplifies to
\be
\begin{split}
\int\frac{\rmd^3\Delta}{(2\pi)^2} \,
\frac{\delta\left(\Delta_\mu \ell^\mu\right)}{\sqrt{4EE'}}
\left[ 
 \frac{A(t)}{m} \left(\ell\cdot P\right)^2 \frac{2\left(P_0^2 - \vec{P}^2 + P_0m\right)}{\cal N}
 +
\frac{J(t)}{m}\, \ell_\mu P^\mu\,
\frac{\ell_\nu P^\nu\, \Delta^2 + m\Delta^2}{\cal N}
\right]~.
\end{split}
\ee
This is required to be $\ge0$ for any $\vec P$ (and any
$\vec{n}$), with $\vec{n}^2=1$.  For $\vec P=0$ specifically it simplifies considerably
since $E=E'$, thus $\Delta^0=0, P^0=E, {\cal N}=E+m$:
\be
\begin{split}
\int\frac{\rmd^3\Delta}{(2\pi)^2} \,
{\delta\left(\vec{n}\cdot \vec\Delta\right)}
\left[ 
 \frac{A(t)}{m} (P^0)^2
 +
\frac{J(t)}{2m}\,\Delta^2
\right] &=
\int\frac{\rmd^2\Delta_\perp}{(2\pi)^2} \,
\left[ 
 \left(A(-\vec\Delta_\perp^2)-2J(-\vec\Delta_\perp^2) \right)
 \frac{\vec\Delta_\perp^2}{4m}+ mA(-\vec\Delta_\perp^2) 
 \right]~.
\end{split}
\ee
Recall that $A-2J \sim 1/\Delta_\perp^6$ asymptotically in QCD so the
integral is convergent.
\\

We now turn to a spin-0 pion. The matrix elements of its EMT are
\be
\begin{split}
\left<\vec P + \frac{\vec\Delta}{2}\left|\, \hat T^{00}(0) \,\right| \vec P -\frac{\vec\Delta}{2}\right>
&= 2 (P^0)^2 A(t) + \frac{1}{2} \vec\Delta^2 D(t)~,\\
\left<\vec P + \frac{\vec\Delta}{2} \left|\, \hat T^{0i}(0) \,\right| \vec P -\frac{\vec\Delta}{2}\right>
&= 2 P^0 P^i A(t) + \frac{1}{2} \Delta^0\Delta^i D(t)~,\\
\left<\vec P + \frac{\vec\Delta}{2} \left|\, \hat T^{ij}(0) \,\right| \vec P -\frac{\vec\Delta}{2}\right>
&= 2 P^i P^j A(t) + \frac{1}{2} \left(\Delta^i\Delta^j+\delta^{ij}\Delta^2\right) D(t)~,
\end{split}
\ee
These are the expressions to be used in eq.~(\ref{eq:T_munu(x0,x)_Appendix}) for 
$T_{\mu\nu}(x^0,\vec x, \vec P)$.

The GFFs of a spin 0 particle have monopole form,
\be
A(t) = \frac{\alpha_{Aq}}{1 -t/\Lambda_{Aq}^2} + \frac{\alpha_{Ag}}{1 -t/\Lambda_{Ag}^2}~,
\ee
and similar for $D(t)$.
The monopole masses $\Lambda_{Aq}, \Lambda_{Ag}$ in the quark and gluon contributions
have been determined on the lattice in ref.~\cite{Hackett:2023nkr}, again up to ${-t}=2$~GeV$^2$.
In the asymptotic regime the pion GFFs from perturbative QCD
have been found to scale $\sim 1/t$~\cite{Tanaka:2018wea,Tong:2021ctu}.
\\

From eq.~(\ref{eq:T_munu(x0,x)_Appendix}) we obtain the components of the Wigner
transform; for simplicity, we again choose the Breit frame.
We then have
\be \label{eq:pion-T_00(x)-fixed}
\begin{split}
T^{00}(\vec x) &= \frac{[\Delta^2 A(-\Delta^2)]_\infty + [\Delta^2 D(-\Delta^2)]_\infty}{4\pi^2\, r^2} 
\\
&  +
\int\frac{\rmd^3\Delta}{(2\pi)^3}\, \, e^{-i\vec\Delta \cdot \vec x}\,
\left[E A(t) + \frac{1}{4E} \vec\Delta^2 D(t) - \frac{[\Delta^2 A(-\Delta^2)]_\infty}{2\Delta}
- \frac{[\Delta^2 D(-\Delta^2)]_\infty}{2\Delta}\right]~.
\end{split}
\ee
Here, $[\Delta^2 A(-\Delta^2)]_\infty = \lim_{\Delta\to\infty}\, \Delta^2 A(-\Delta^2)$
with $\Delta = |\vec\Delta|$.
In the Breit frame, $T^{0i}=0$.
Lastly,
\be \label{eq:pion-T_ij(x)-fixed}
\begin{split}
T^{ij}(\vec x) &=
- [\Delta^2 D(-\Delta^2)]_\infty \frac{x^i x^j}{2\pi^2 r^4}
+ \int\frac{\rmd^3\Delta}{(2\pi)^3}\, \, e^{-i\vec\Delta \cdot \vec x}\,
\left(\Delta^i\Delta^j - \delta^{ij} \vec\Delta^2\right)
\left[ \frac{D(t)}{4E}
- \frac{[\Delta^2 D(-\Delta^2)]_\infty}{2\Delta^3}
\right]~.
\end{split}
\ee
\\

We now write $x^\mu(\zeta) = x_0^\mu + \zeta L\ell^\mu = (\zeta ,\vec x_0 + \zeta L\hat n)$, so $\rmd x^\mu/\rmd\zeta = L \ell^\mu = L (1,\hat n)$ where $L$ denotes an arbitrary length scale.
Also, $\vec{x}^2 = (\vec x_0 + \zeta L\hat n)^2 = {x_0}^2 + 2 \zeta L \vec x_0 \cdot\hat{n}
+ L^2 \zeta^2$.
Note that here, for the pion, a shift of the geodesic from the center is required since
some components of $T_{ij}(\vec x)$ diverge as $|\vec x| = r \to 0$.
The ANEC integral for the pion is given by
\be
\begin{split}
\int\limits_{-\infty}^\infty\rmd\zeta \, \ell^\mu \ell^\nu \, T_{\mu\nu}(\vec x) &= 
\int\limits_{-\infty}^\infty\rmd\zeta \, \left[T_{00}(\vec x) + \hat{n}^i\hat{n}^j T_{ij}(\vec x)
\right]~.
\end{split}
\ee
The first term with $T_{00}$ gives
\be
\begin{split}
&
\frac{[\Delta^2 A(-\Delta^2)]_\infty + [\Delta^2 D(-\Delta^2)]_\infty}{4\pi L \sqrt{{x_0}^2-(\vec x_0 \cdot\hat{n})^2}}\\
&  +
\int\frac{\rmd^2\Delta_\perp}{(2\pi)^2 L}\, e^{-i \vec{\Delta}_\perp\cdot\vec{x}_{0\perp}}
\left[E A(-\Delta_\perp^2) + \frac{\Delta_\perp^2}{4E}  D(-\Delta_\perp^2) - 
\frac{[\Delta^2 A(-\Delta^2)]_\infty}{2\Delta_\perp}
- \frac{[\Delta^2 D(-\Delta^2)]_\infty}{2\Delta_\perp}\right]
~,
\end{split}
\ee
while the term involving $T_{ij}$ gives
\be
\begin{split}
 & -  \frac{[\Delta^2 D(-\Delta^2)]_\infty}{4\pi L \sqrt{{x_0}^2-(\vec x_0 \cdot\hat{n})^2}}
 - \int\frac{\rmd^2\Delta_\perp}{(2\pi)^2 L}\, \, e^{-i\vec\Delta_\perp \cdot \vec{x}_{0\perp}}\,
\vec\Delta_\perp^2 \left[ \frac{D(-\Delta_\perp^2)}{4E}
- \frac{[\Delta^2 D(-\Delta^2)]_\infty}{2\Delta_\perp^3}
\right]
\end{split}
\ee
Adding both contributions,
\be
\begin{split}
\mathrm{ANEC}_\pi
&=
\frac{[\Delta^2 A(-\Delta^2)]_\infty}{4\pi L \, |\vec{x}_{0\perp}|}
 +
\int\frac{\rmd^2\Delta_\perp}{(2\pi)^2 L}\, e^{-i \vec{\Delta}_\perp\cdot\vec{x}_{0\perp}}
\left[E A(-\Delta_\perp^2) - 
\frac{[\Delta^2 A(-\Delta^2)]_\infty}{2\Delta_\perp}\right]
~.
\end{split}
\ee
This is the final result for ANEC for the pion. For $\Delta_\perp \to \infty$ the
bracket under the integral is $\sim 1/\Delta_\perp^3$, hence the integral is finite. 
The first term is always positive and
letting $|\vec{x}_{0\perp}|$ become arbitrarily small,
ANEC$_\pi$ exhibits a power-divergence.
So, in this limit ANEC$_\pi \ge 0$ should be satisfied.

On the other hand, we have found numerically that a monopole form factor
$A(-\Delta^2)=1/(1+\Delta^2/\Lambda^2)$ parameterized in terms of
a {\em single} monopole mass across the entire range of momentum transfer,
breaks ANEC$_\pi \ge 0$ (for $|\vec{x}_{0\perp}|\, \Lambda \gtrsim 1$)
when the pion mass $m_\pi / \Lambda$ is sufficiently small.
This suggests that
the Lattice monopole fit mentioned above will not hold to high momentum transfer; 
the $A-$GFF will change for higher $\Delta$ before it
settles, asymptotically, to yet again a monopole form.
Indeed, such behavior has been observed in some lattice QCD computations of
$A(t)$~\cite{Gao:2022vyh}, however, without relating it to ANEC.

\bibliography{references}

\end{document}